\documentclass[conference, 10pt]{IEEEtran}

\usepackage{amsmath,epsfig}
\usepackage{mathptmx}
\usepackage[ruled,vlined]{algorithm2e}
\usepackage{float}
\usepackage{amsmath}
\usepackage{graphicx}
\usepackage{color}
\usepackage{placeins}
\usepackage{float}
\usepackage{tabularx,colortbl}

\let\labelindent\relax
\usepackage{enumitem}
\usepackage{url}
\usepackage{hyperref}

\begin{document}

\onecolumn 

\begin{description}[labelindent=1cm,leftmargin=3cm,style=multiline]

\item[\textbf{Citation}]{D. Temel and G. AlRegib, "Image quality assessment and color difference," 2014 IEEE Global Conference on Signal and Information Processing (GlobalSIP), Atlanta, GA, 2014, pp. 970-974.
} \\

\item[\textbf{DOI}]{\url{https://doi.org/10.1109/GlobalSIP.2014.7032265}} \\

\item[\textbf{Review}]{Date added to IEEE Xplore: February 9, 2015} \\

\item[\textbf{Poster}]{\url{https://ghassanalregib.com/publications/}} \\

\item[\textbf{Bib}] {
@INPROCEEDINGS\{Temel2014\_GlobalSIP\_IQA,\\ 
author=\{D. Temel and G. AlRegib\},\\ 
booktitle=\{2014 IEEE Global Conference on Signal and Information Processing (GlobalSIP)\},\\ 
title=\{Image quality assessment and color difference\},\\ 
year=\{2014\},\\
pages=\{970-974\},\\  
doi=\{10.1109/GlobalSIP.2014.7032265\},\\ 
month=\{Dec\},\}\\
} \\

\item[\textbf{Copyright}]{\textcopyright 2014 IEEE. Personal use of this material is permitted. Permission from IEEE must be obtained for all other uses, in any current or future media, including reprinting/republishing this material for advertising or promotional purposes,
creating new collective works, for resale or redistribution to servers or lists, or reuse of any copyrighted component
of this work in other works. } \\

\item[\textbf{Contact}]{\href{mailto:alregib@gatech.edu}{alregib@gatech.edu}~~~~~~~\url{https://ghassanalregib.com/}\\ \href{mailto:dcantemel@gmail.com}{dcantemel@gmail.com}~~~~~~~\url{http://cantemel.com/}}
\end{description} 

\thispagestyle{empty}
\newpage
\clearpage

\twocolumn

%
\title{Image Quality Assessment and Color Difference}

\author{\IEEEauthorblockN{Dogancan Temel and Ghassan AlRegib}
\IEEEauthorblockA{Center for Signal and Information Processing (CSIP)\\
Electrical and Computer Engineering\\
Georgia Institute of Technology\\
Atlanta, Georgia, USA\\
\{cantemel,alregib\}@gatech.edu\\}
}

\maketitle

\begin{abstract}
An average healthy person does not perceive the world in just black and white. Moreover, the perceived world is not composed of pixels and through vision humans perceive structures. However, the acquisition and display systems discretize the world. Therefore, we need to consider pixels, structures and colors to model the quality of experience. Quality assessment methods use the pixel-wise and structural metrics whereas color science approaches use the patch-based color differences. In this work, we combine these approaches by extending CIEDE2000 formula with perceptual color difference to assess image quality. We examine how perceptual color difference-based metric  (\verb"PCDM") performs compared to PSNR, CIEDE2000, SSIM, MS-SSIM and CW-SSIM on the LIVE database. In terms of linear correlation, \verb"PCDM" obtains compatible results under white  noise (97.9\%), Jpeg (95.9\%) and Jp2k (95.6\%) with an overall correlation of 92.7\%. We also show that \verb"PCDM" captures color-based artifacts that can not be captured by structure-based metrics.

\end{abstract}

\begin{IEEEkeywords} color-difference, perceptual quality, objective quality metrics, color artifacts \end{IEEEkeywords}

\IEEEpeerreviewmaketitle

\section{Introduction}
\label{sec:intro}
    \vspace{-0.1cm}
The phrase "quality of experience" in the image processing literature indicates the perceived quality of images. Therefore, perception matters as much as the fidelity for consumer electronics applications. Images are commonly analyzed in terms of pixels and structures in the image processing literature. However, color science literature mostly focus on large patches that are functions of visual fields. In our view, in order to model the full quality of experience, we need to consider the literature of both image processing and color science. The proposed approach contributes to the literature by utilizing the color label differences in the quality of experience estimation.

Objective quality metrics are used in the image processing literature to estimate the quality of experience or to quantify distortions. Pixel-wise fidelity metrics focus on the exact differences between pixels of the images. As an example, root-mean-square error and peak signal-to-noise ratio (PSNR) are commonly used in the literature because of their simplicity. Instead of calculating the  pixel-wise fidelity, structural fidelity of  images are also used to estimate the quality. SSIM  calculates the local statistics of images over a single scale whereas MS-SSIM  follows a multi-scale approach to calculate SSIM over different resolutions using Laplacian pyramid as described in \cite{bovik2003}. CW-SSIM \cite{sampat2009} also follows a multi-scale approach but instead of calculating the local statistics in the spatial domain, wavelet coefficients are used. Most of these quality metrics use the luminance components or grayscale images and neglect the color channels. 

In contrast to the luminance-based image quality measures, color information is commonly used in the color science literature to detect the differences between similar color tones  \cite{wyszecki1982}.
The International Commission on Illumination (CIE) is  responsible for the international coordination of lighting related technical standards including color difference. CIEDE2000 color difference equation was developed by the CIE technical committee and it is one of the state of the art metrics in the color science literature as described in  \cite{luo2001}, \cite{dalal2005}. Color differences and similarities can  be used as  descriptors of the images. The authors in \cite{weijer2009} propose learning color names from real-world images, which can be used for object recognition and image classification as described in \cite{weijer2008}. Color naming descriptors are also used in image classification in terms of the aesthetics quality of the images as explained in  \cite{temel2014}. Moreover, color naming descriptors are  used in \cite{pele2012} to 
perform color-based edge detection.

\begin{figure*}[ht!]
\centering
\includegraphics[width=0.60\linewidth]{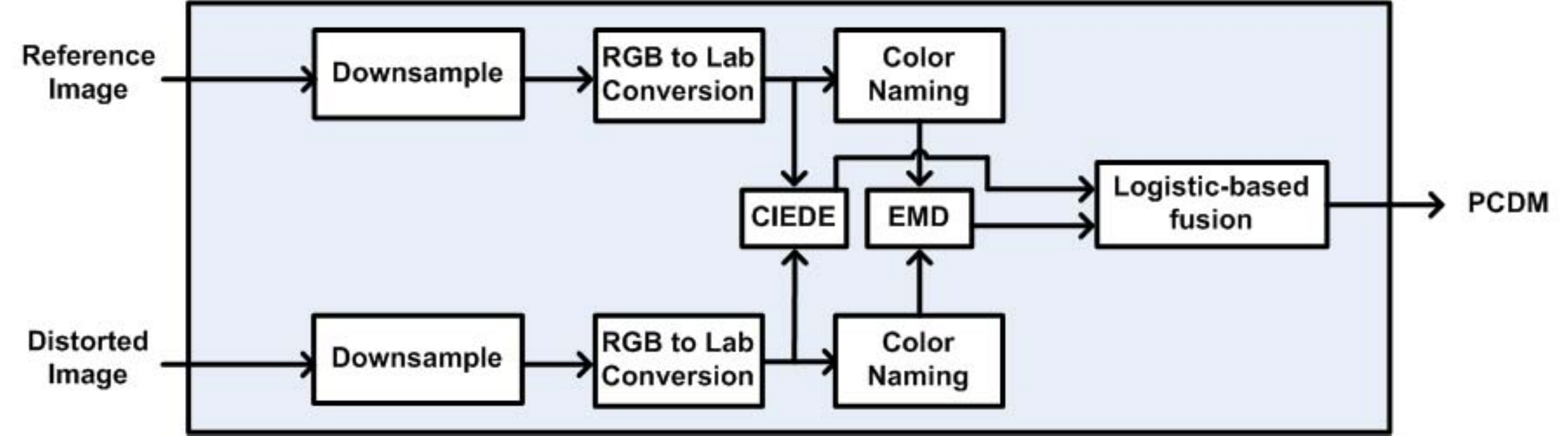}
  \vspace{-0.2cm}
\caption{Perceptual Color Difference-based Metric Pipeline}
\label{fig:pcdmPipeline}
  \vspace{-0.8cm}
\end{figure*}

Color difference formulas are commonly used in tone matching for color reproduction.  The authors in \cite{zhang1997} use a color difference-based metric to predict texture visibility of printer halftone patterns. The way color difference is used in \cite{zhang1997} can be considered as a transition in the application field of color difference equations from basic tone matching to textured image comparisons. In \cite{johnson2003}, the authors discuss the connections between image quality, difference and appearance. Even color image quality is discussed in these approaches, authors consider the problem from the point of color science and leave the discussion limited without fully studying the objective quality metrics and their performance under different kinds of distortions such as compression and communication errors. The authors in \cite{yang2012} take the difference equations one step further and describe a calibration process for the color difference equations under experimental conditions as well as the usage of CIEDE2000 as an image quality metric. However, a very fundamental characteristic of the difference equations is overlooked. In principle, CIEDE2000 is designed for tone matching between similar colors that are bounded by, at most, medium differences and not utilized for significant tone differences.

In this paper, we augment the range of CIEDE2000 formula with perceptual color difference as in \cite{pele2012} to generalize the method proposed by the authors in \cite{yang2012}. In Section \ref{sec:main}, we describe the main blocks in the proposed quality metric pipeline. We discuss the experimental setup, results and observations in Section \ref{sec:results} and conclude our discussion in Section \ref{sec:conclusion}.

\section{Color-based Image Quality Assessment}
\label{sec:main}
The pipeline of the proposed image-quality assessment method is given is Fig. \ref{fig:pcdmPipeline}. Since human visual system is less sensitive to the color compared to the structure, color-based image quality can be calculated over the smoothed version of the image. Images are downsampled using the defaults of bicubic interpolation and anti-aliasing. Sampling rate is set to $0.05$ after exhaustively simulating the range from 0.02 to 0.25. The effect of downsampling is explained in Section \ref{subsec:results}. After downsampling, the \texttt{RGB} images are converted to the \texttt{Lab} domain to represent pixels in a perceptually correlated color space. Color descriptors are calculated for each pixel as it is described in Section \ref{subsec:main_colornaming}. We explain the usage of the  Earth Mover's Distance (EMD), the CIEDE2000 formula and the logistic-based fusion function in Section \ref{subsec:main_colorDifference}.

\subsection{Color Naming}
\label{subsec:main_colornaming}
\label{colorNaming}
Linguistic color names are used to label image pixels with perceived color classes. The authors in \cite{weijer2009} introduced color naming as  a $11-D$ image descriptor. Each dimension in the descriptor corresponds to the probability of that pixel to be perceived as one of the $11$ basic colors. These basic color names are: black, blue, brown, grey, green, orange, pink, purple, red, white and yellow. In order to determine the pixel values for the finite color vocabulary, Google Image is used to obtain the training set, which also includes  the wrongly labeled images. Color names are learned from the noisy data using the variants of the probabilistic latent semantic analysis model as explained in \cite{weijer2009}. In addition to $11-D$ color descriptors, we have also experimented $25-D$ and $50-D$ descriptors that span a wider color range. However, we have not observed a significant increase in the estimation accuracy.

\subsection{Color Difference}
\label{subsec:main_colorDifference}

The CIEDE2000 color difference equation is designed to calculate the difference between similar colors with low level variations. In the proposed metric, we keep the display and viewing related parameters constant to make the metric independent of acquisition and display configurations. In order to limit the calculation of CIEDE2000 to low-level differences, a threshold is set for color difference values and the result is divided by the same threshold to normalize the difference to be between $0$ and $1$. S-CIELab \cite{wandell1997}  color difference equation is not used in the proposed metric to eliminate parameter tuning including but not limited to spatial and color calibration.

\begin{figure}[htbp!]
\begin{center}
\noindent
  \includegraphics[width=0.75\linewidth]{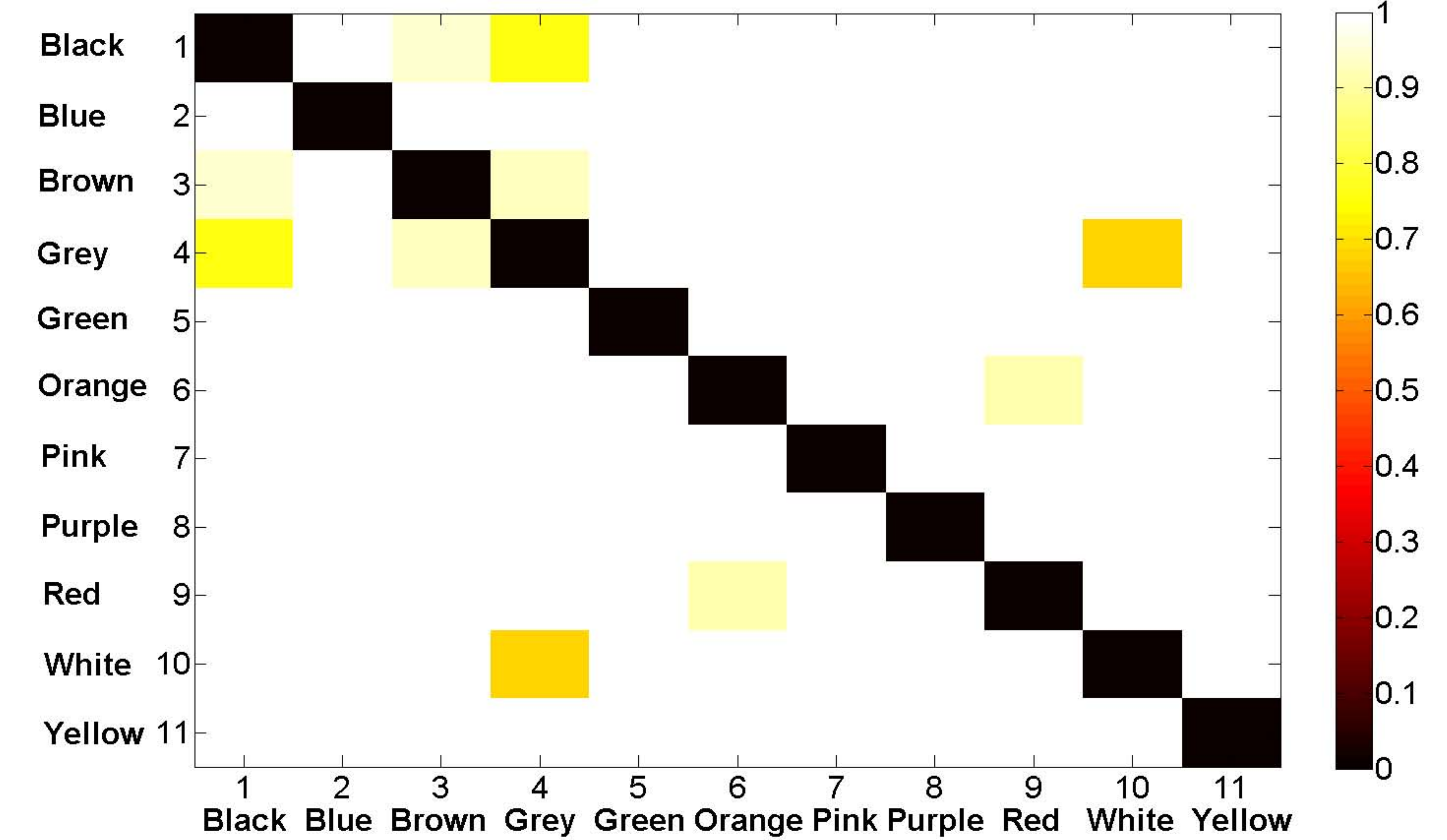}
    \vspace{-0.2cm}
  \caption{Perceived distance between basic color categories.}
  \label{ColorMat}
    \vspace{-0.5cm}
\end{center}
\end{figure}

The Earth Mover's Distance (EMD) is designed to calculate the difference between two distributions. The basic idea behind EMD is to calculate the minimal cost that is required to transfer one distribution into the other \cite{rubner2000}. In \cite{pele2012}, EMD is calculated between two $11-D$ color naming descriptors where the flow between each color label probability is calculated to obtain the cost. Instead of using the uniform distance, the flow between color labels is scaled according to the perceived color distance. We can calculate this distance by using the joint distribution of basic color terms in the \texttt{Lab} color cube as explained in \cite{weijer2009}. The visualization of the perceived distance is depicted in Fig. \ref{ColorMat}.

    \vspace{-0.4cm}

 EMD considers all flow scenarios from the source  to the target color descriptor to minimize the total cost. We can formulate the EMD expression as in  Eq. \ref{eq:EMD} where $i$ is the index of the color label in the reference color descriptor and $j$ is the index of the compared color descriptor. Flow from the $i_{th}$ color probability in the reference to the $j_{th}$ color probability in the compared descriptor is represented by $f_{ij}$ and the perceived distance between the color terms in the dictionary is shown with $d_{ij}$.

 \vspace{-0.4cm}
\begin{equation}
\label{eq:EMD}
EMD=\underset{f_{i,j}}{min}\left \{ \sum_{i=1}^{11} \sum_{j=1}^{11}d_{i,j}f_{i,j} \right \}
\end{equation}

The authors in \cite{pele2012} combined CIEDE2000 with EMD \cite{rubner2000} using a logistic function to obtain a perceptually correlated difference function as formulated in Eq. \ref{eq:coldist}. The difference function is used to detect edges. $S_i$ is the \texttt{Lab} value and $P_i$ is the color naming descriptor corresponding to a single pixel of the image indexed with $i$. $alpha$ is set to $0.5$ to equivalently combine CIEDE2000 and EMD and and $z$ is set to $10$ as in \cite{pele2012}. The parameters in the logistic-bsaed fusion are selected independent from the tested image database.

  \vspace{-0.2cm}
\begin{equation}
\label{eq:coldist}
\frac{1}{1+e^{-z(((\alpha)CIEDE(S_1,S_2)+(1-\alpha)EMD(P_1,P_2))-\frac{1}{2})}}
\end{equation}

In the proposed pipeline, we asses the perceived quality of the images by extending the range of the color difference equation.  CIEDE2000 is unreliable when the color difference is more than 7 \texttt{CIELAB}. Therefore, we threshold the color difference equation for high level differences and just use the transportation distance between the color descriptors that are inherently used for images significantly different from each other. The scope of the proposed metric is to introduce the perceptual color differences into image quality estimation. As a future work, other distance metrics can also be used to quantify the difference between color descriptors.

%
%
%

\section{Experimental Evaluation}

\label{sec:results}

\subsection{Experimental Setup}
\label{subsec:main_dataset}
In this paper, we use the release $2$ version of the LIVE image database for the validation of the proposed metric. The resolution of most of the images in the database is $768x512$ and images are interpolated (bicubic) to $1024x768$ for subjective tests. There are $29$ reference images and $779$ distorted images. JPEG, JPEG2000 (Jp2k), White Noise (Wn), Gaussian Blur (Gblur) and Simulated Fast Fading Rayleigh Channel errors (FF) are the main sources of degradation in the image database. A more detailed information related to LIVE image database can be found in  \cite{sheikh2006}. As it is explained in \cite{sheikh2006}, the non-linear regression formulated in Eq. \ref{eq:nonlinreg} is applied to each objective quality metric to make a fair comparison. 

    \vspace{-0.4cm}
\begin{equation}
\label{eq:nonlinreg}
S=\beta_1 \left ( \frac{1}{1}-\frac{1}{2+exp(\beta_2(S_0 -\beta_3 ))} \right )+\beta_4 S_0 +\beta_5
\end{equation}

 \verb"PCDM" is used to refer to the proposed perceptual color difference-based metric in the rest of the paper. We compare the performance of \verb"PCDM" with PSNR, CIEDE2000 SSIM, MS-SSIM and CW-SSIM, which were described in Section \ref{sec:intro}.




\subsection{Results}
\label{subsec:results}

\begin{figure}[htbp!]
\begin{minipage}[b]{0.48\linewidth}
  \centering
\includegraphics[width=0.75\linewidth, trim= 0mm 74mm 0mm 65mm]{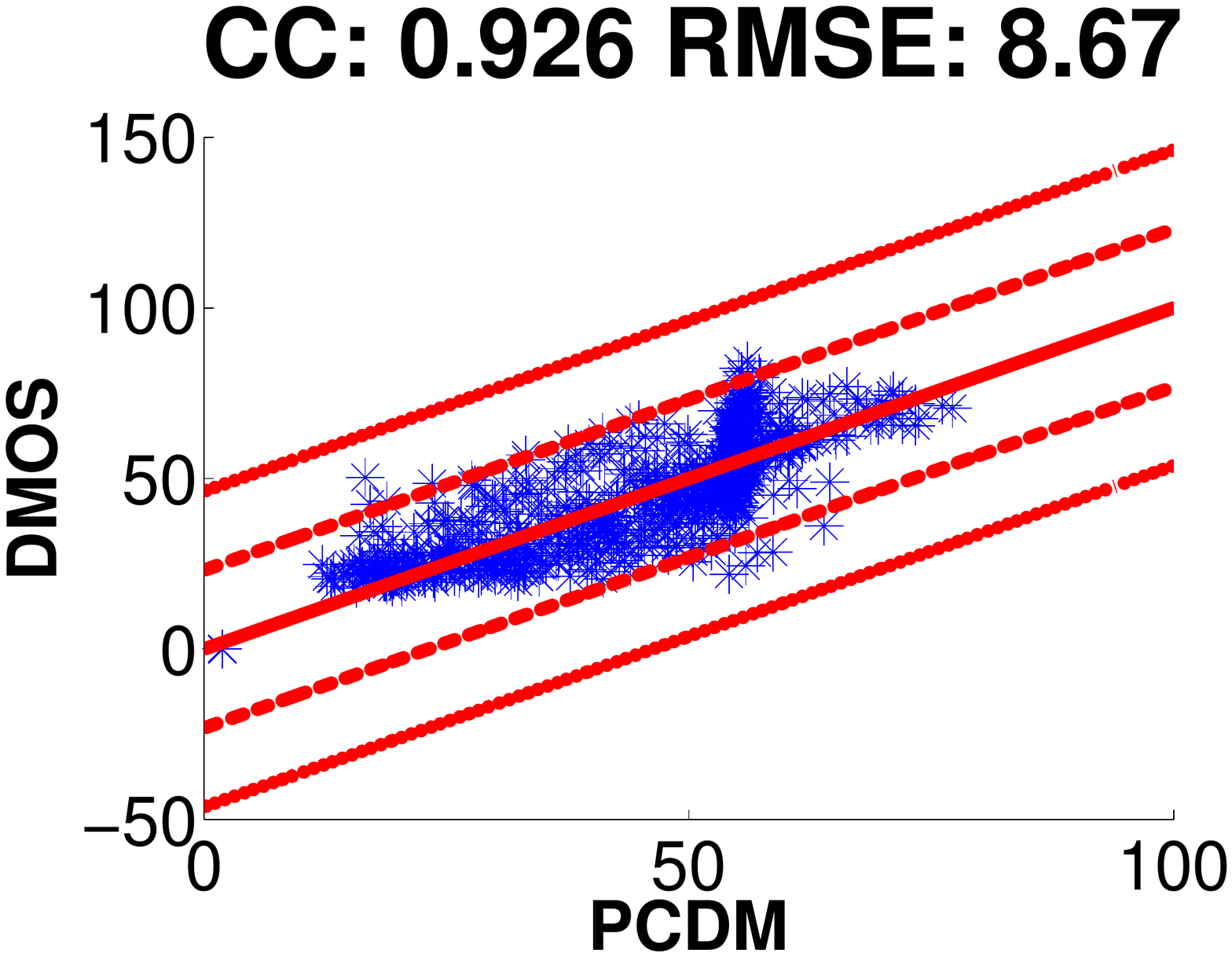}
  \vspace{0.08cm}
  \centerline{\footnotesize{(a) Full Image Set}}
\end{minipage}
  \vspace{0.2cm}
\hfill
\begin{minipage}[b]{.48\linewidth}
  \centering
\includegraphics[width=0.75\linewidth, trim= 0mm 74mm 0mm 65mm]{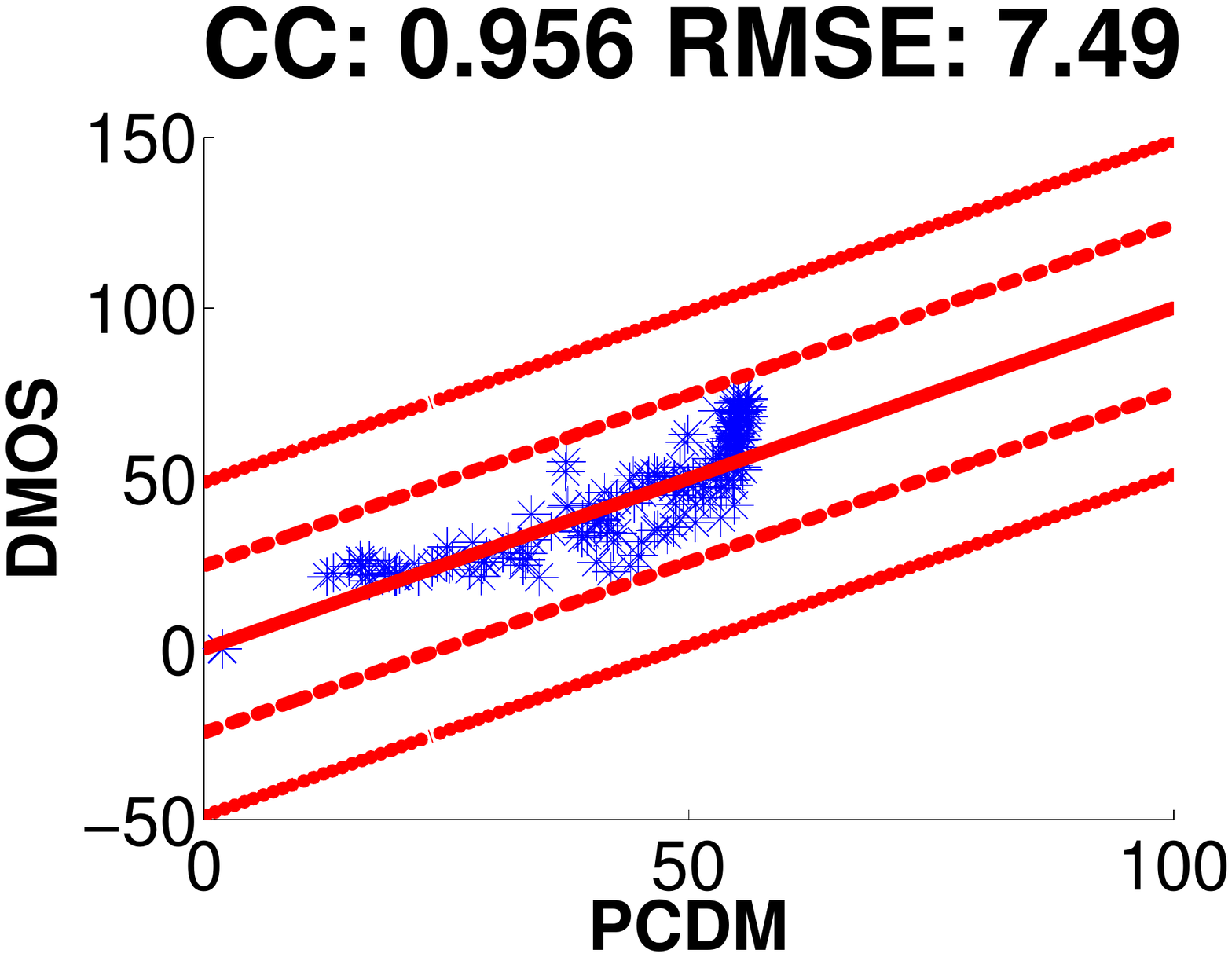}
  \vspace{0.08cm}
  \centerline{\footnotesize{(b) Jp2k Compression}}
\end{minipage}

\vspace{-.2cm}

\begin{minipage}[b]{.48\linewidth}
  \centering
\includegraphics[width=0.75\linewidth, trim= 0mm 74mm 0mm 65mm]{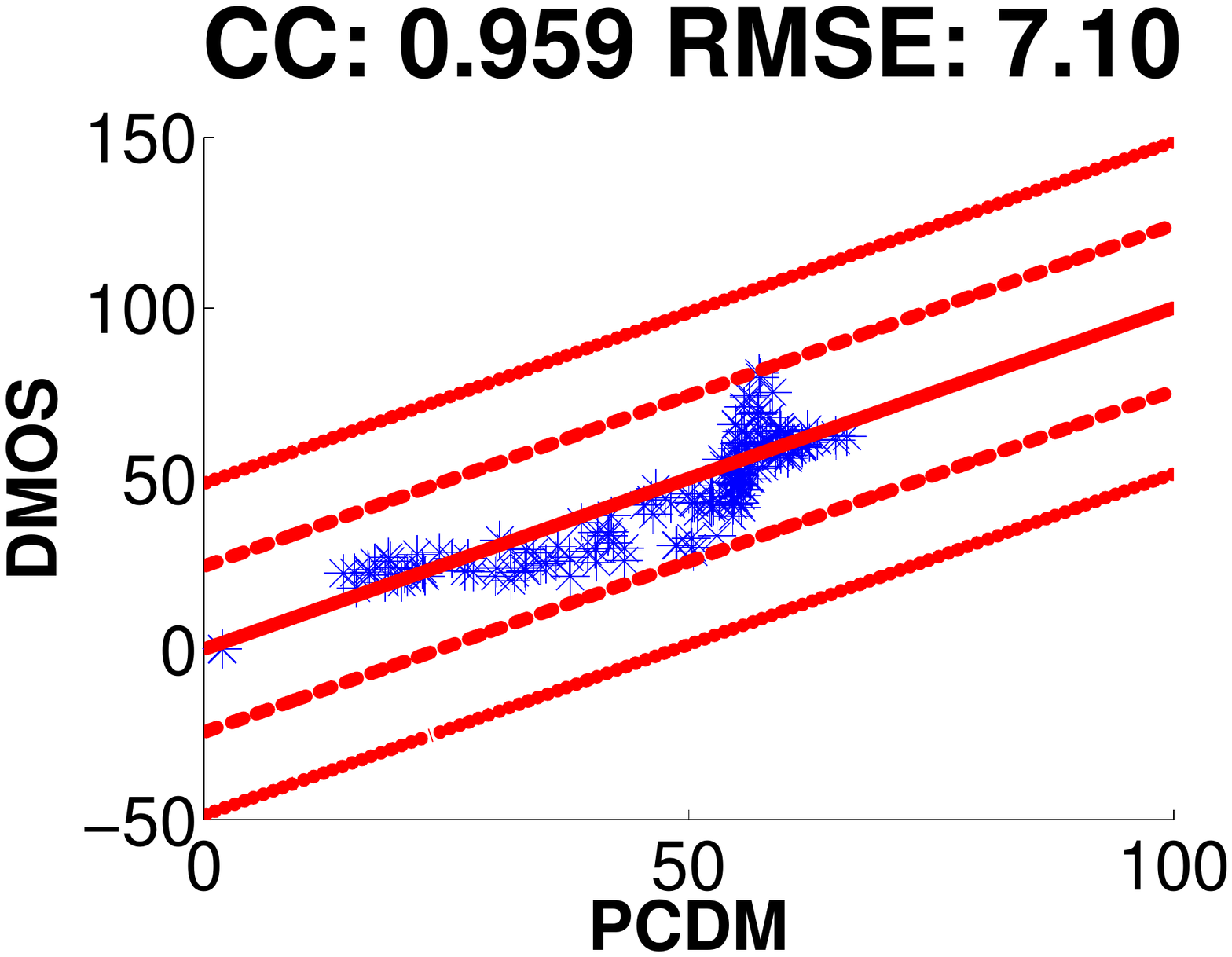}
  \vspace{0.08 cm}
  \centerline{\footnotesize{(c) Jpeg Compression } }
\end{minipage}
  \vspace{0.3cm}
\hfill
\begin{minipage}[b]{0.48\linewidth}
  \centering
\includegraphics[width=0.75\linewidth, trim= 0mm 74mm 0mm 65mm]{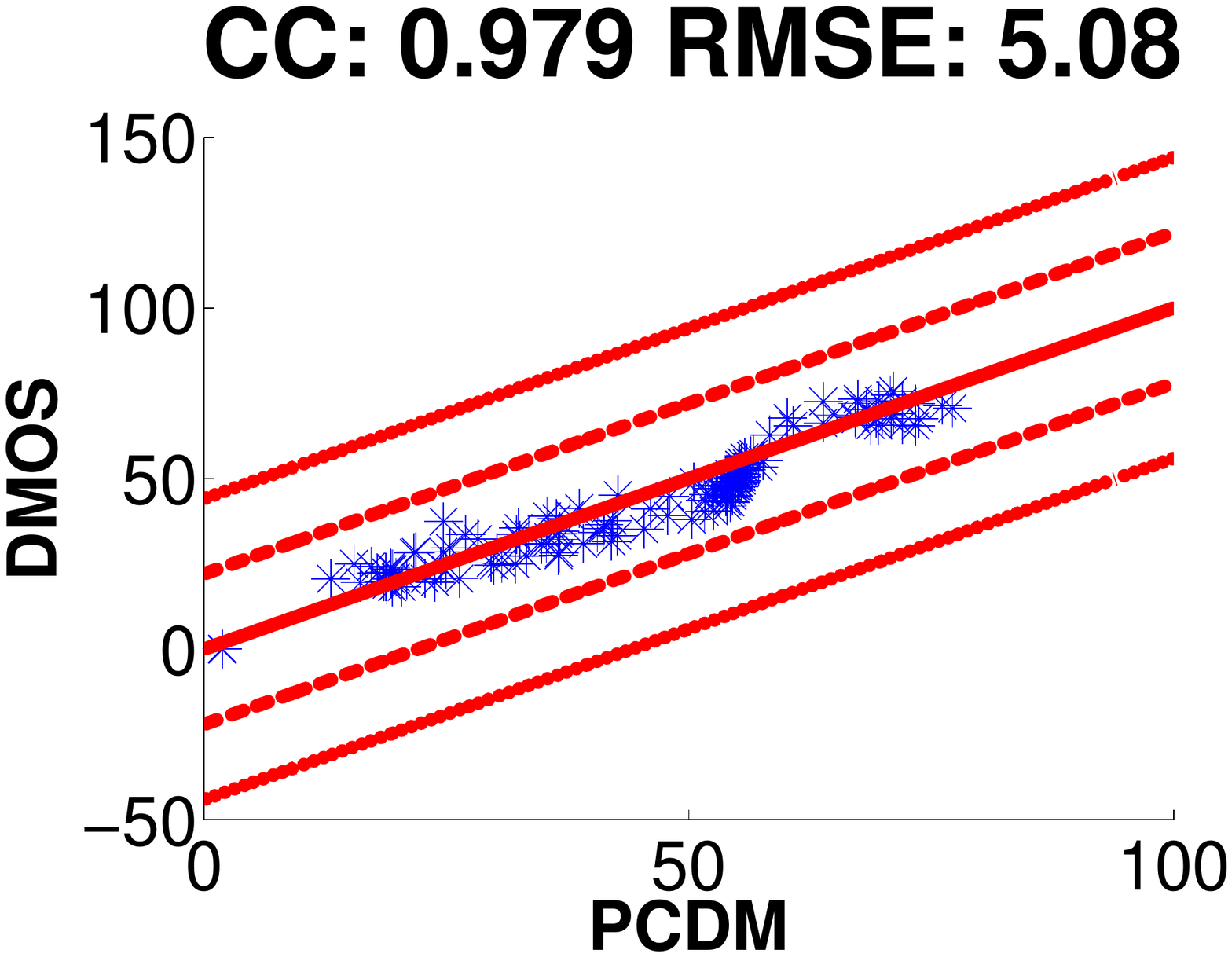}
  \vspace{0.08 cm}
  \centerline{\footnotesize{(d) White Noise}}
\end{minipage}
\vspace{-.3cm}
\begin{minipage}[b]{.48\linewidth}
  \centering
\includegraphics[width=0.75\linewidth, trim= 0mm 74mm 0mm 74mm]{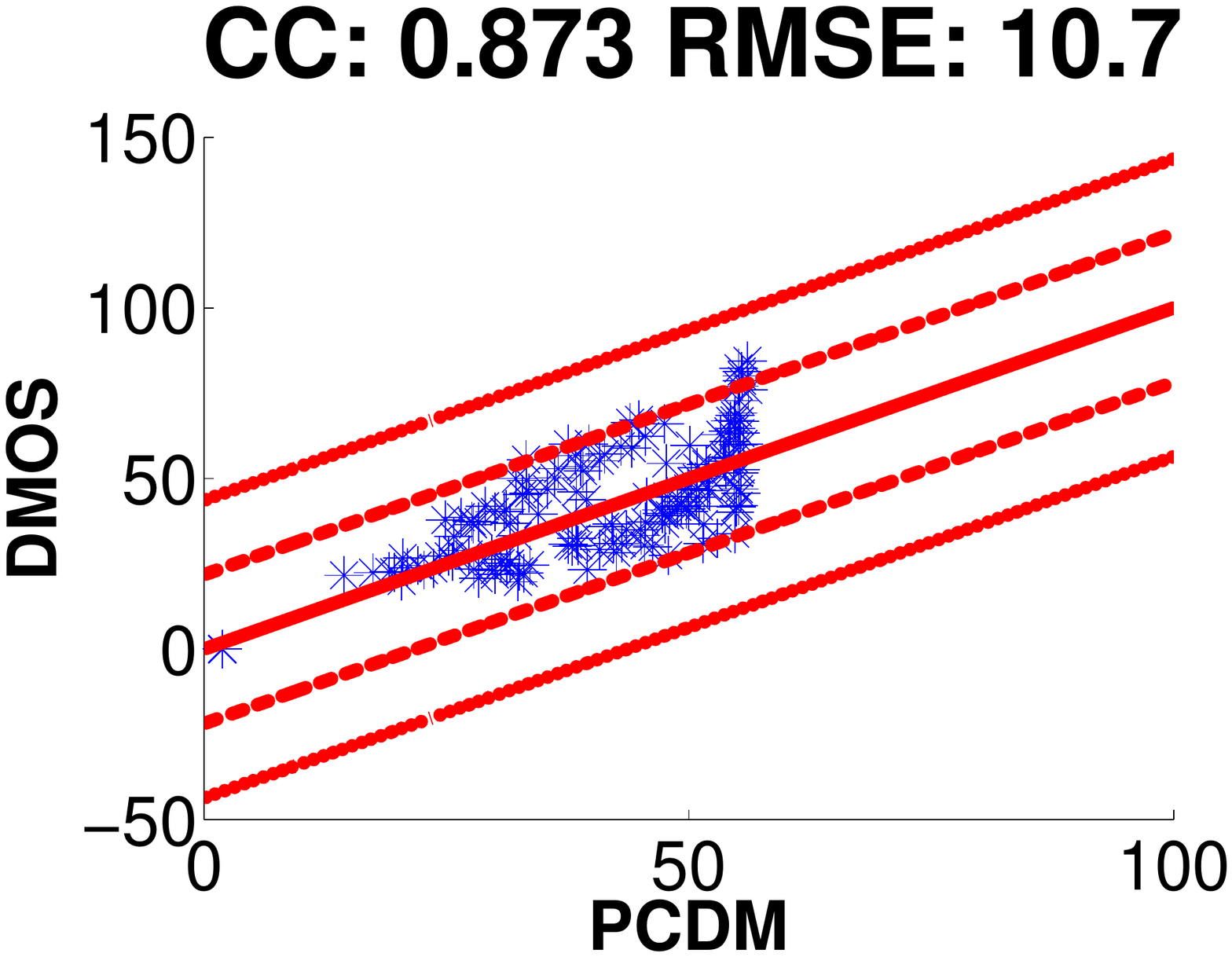}
  \vspace{0.08cm}
  \centerline{\footnotesize{(e) Gaussian Blur } }
\end{minipage}
\hfill
\begin{minipage}[b]{0.48\linewidth}
  \centering
\includegraphics[width=0.75\linewidth, trim= 0mm 74mm 0mm 74mm]{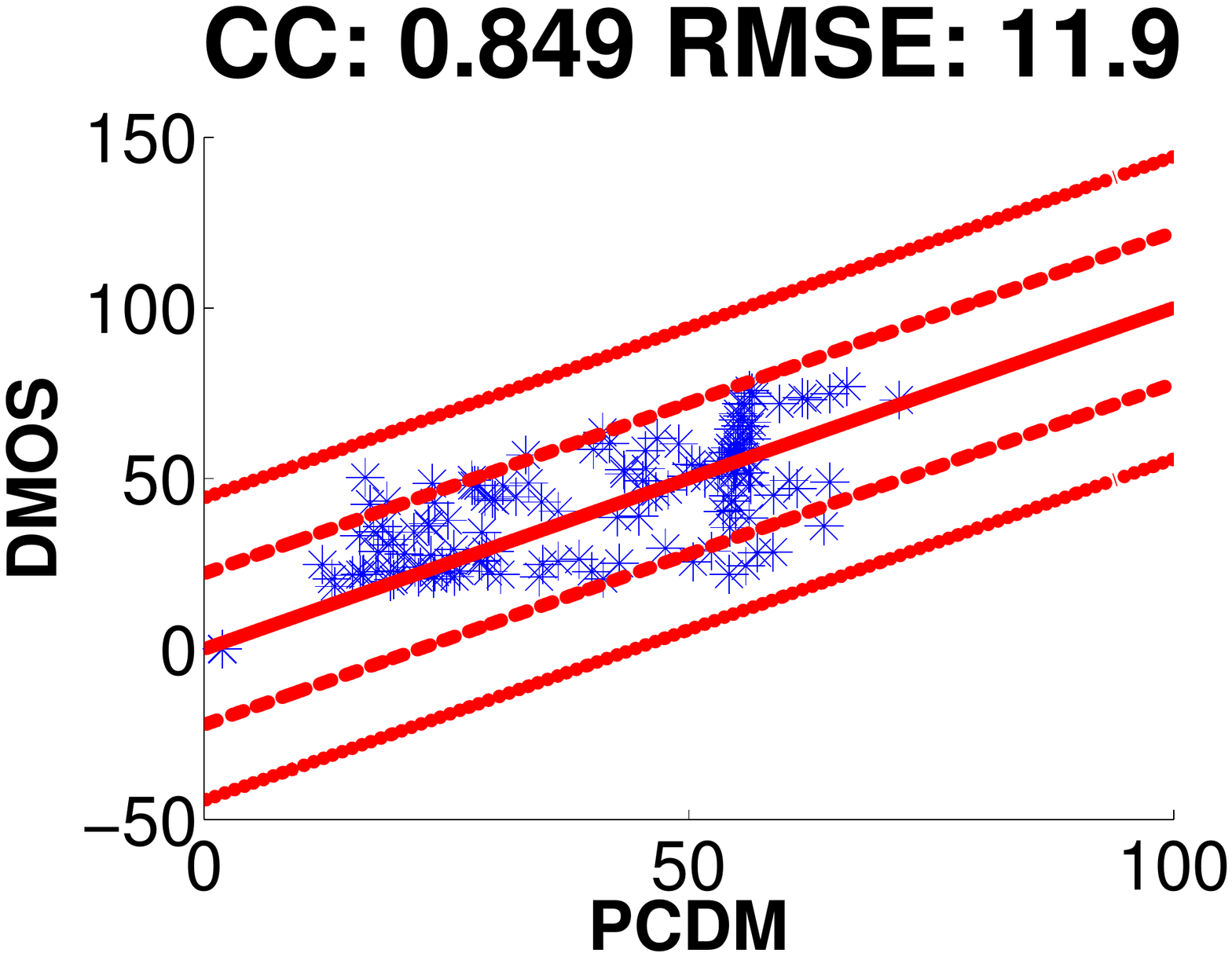}
  \vspace{0.08 cm}
  \centerline{\footnotesize{(f) Rayleigh Fastfading}}
\end{minipage}

  \vspace{-0.1 cm}
\caption{Scatter plots of DMOS versus PCDM after non-linear regression.}
\label{fig:scatterplots}
  \vspace{-0.4 cm}

\end{figure}

The scatter plots of  \verb"PCDM" under different distortion types are given in Fig. \ref{fig:scatterplots}. Solid lines correspond to the ideal scenario where quality estimation is equivalent to the average subjective scores. Dashed lines are located one standard deviation away from the solid line and dotted lines are two standard deviations away. Linear correlation coefficient (CC) and root-mean-square error (RMSE) are calculated between the  quality metric estimates and the difference mean opinion scores (DMOS) after non-linear regression. High CC values mean that the relation between the estimates and the subjective scores are highly linear. Low RMSE values indicate  the proximity of the estimates to the DMOS values.

\verb"PCDM" accurately estimates the subjective results under White Noise with a CC of $0.979$ and RMSE of $5.08$ as shown in Fig. \ref{fig:scatterplots}(d). In the cases of Jpeg and Jp2k, nearly all of the estimates are in the distance of one standard deviation with CC values around 0.95 and RMSE values ranging from $7.10$ to $7.49$ as given in Fig. \ref{fig:scatterplots}(b)-(c).  However, when the images are distorted with Gaussian Blur and Fast Fading,  some of the \verb"PCDM"-based estimates exceed the one standard deviation boundary with CC values less than 0.9 and RMSE values higher than 10.0 as shown in Fig. \ref{fig:scatterplots}(e)-(f). The performance comparison between \verb"PCDM" and other objective quality metrics are given in Table \ref{tab:metricstats}.

\verb"PCDM", CIEDE2000, SSIM and MS-SSIM are close to each other in terms of accurately estimating the subjective results under Jp2k, Jpeg and Wn distortions. However, both of the color-based metrics perform poorly compared to SSIM and MS-SSIM under FF and Gblur.
\verb"PCDM" performs better than CW-SSIM under all types of distortions and it performs better than PSNR under all of the distortion types except FF. Extending the range of CIEDE2000 with perceptual color difference leads to a decrease of $0.35$ in the RMSE value and an increase of $0.07$ in the CC value.

\begin{table}[htbp]
\begin{center}
\caption{ Performance of the objective quality metrics} \label{tab:metricstats}
\begin{tabular}{|c||c|c|c|c|c|c|}
 \hline
 \textbf{Metrics}  & \textbf{Jp2k}  & \textbf{Jpeg}  &\textbf{Wn} &\textbf{Gblur} &\textbf{FF} &\textbf{All}\\
  \hline
  \multicolumn{7}{|c|}{\textbf{Pearson CC (Linear) }}     \\
  \hline
 \textbf{PSNR} & 0.923 & 0.913 & 0.945 & 0.843 & 0.887 & 0.898\\
 \hline
 \textbf{CIEDE2000} & 0.954 & 0.956 & 0.981 & 0.892 & 0.850 & 0.920\\
 \hline
  \textbf{SSIM} & 0.963 & 0.957 & 0.976 & 0.940 & 0.956 & 0.945\\
 \hline
  \textbf{MS-SSIM} & 0.962 & 0.961 & 0.977 & 0.943 & 0.948 & 0.946\\
 \hline
 \textbf{CW-SSIM} & 0.926 & 0.927 & 0.949 & 0.768 & 0.835 & 0.872\\
 \hline
 \textbf{PCDM} & 0.956 & 0.959 & 0.979 & 0.873 & 0.849 & 0.927\\
 \hline
  \multicolumn{7}{|c|}{\textbf{RMSE }}     \\
  \hline
 \textbf{PSNR} & 9.92 & 10.10 & 8.34 & 11.80 & 10.22 & 10.12\\
 \hline
 \textbf{CIEDE2000} & 7.61 & 7.79 & 5.64 & 11.33 & 11.92 & 9.02\\
 \hline

  \textbf{SSIM} & 7.11 & 7.74 & 8.65 & 7.54 & 6.45 & 7.52\\
 \hline
 \textbf{MS-SSIM} & 7.12 & 7.30 & 8.38 & 7.38 & 7.04 & 7.43\\
 \hline
 \textbf{CW-SSIM} & 9.75 & 9.30 & 9.24 & 14.45 & 13.62 & 10.87\\

 \hline
 \textbf{PCDM} & 7.49 & 7.10 & 5.08 & 10.73 & 11.96 & 8.67\\
 \hline

\end{tabular}
\end{center}
\vspace{-0.2cm}
\end{table}

The authors in \cite{sheikh2006} examine quality versus distortion parameter distributions for the LIVE database. In the case of Fast Fading, data points are distributed all around the scatter plot with a high variation. Because there is not a high correlation between the distortion parameters and DMOS, it is expected that \verb"PCDM" would not be a good model to estimate Fast Fading. However, there is a high correlation between the distortion parameters and DMOS in Gaussian Blur and hence \verb"PCDM" is expected to model the Gaussian Blur distortions.

\begin{figure}[htbp!]
\begin{minipage}[b]{0.48\linewidth}
  \centering
\includegraphics[width=0.60\linewidth, trim= 25mm 90mm 25mm 90mm]{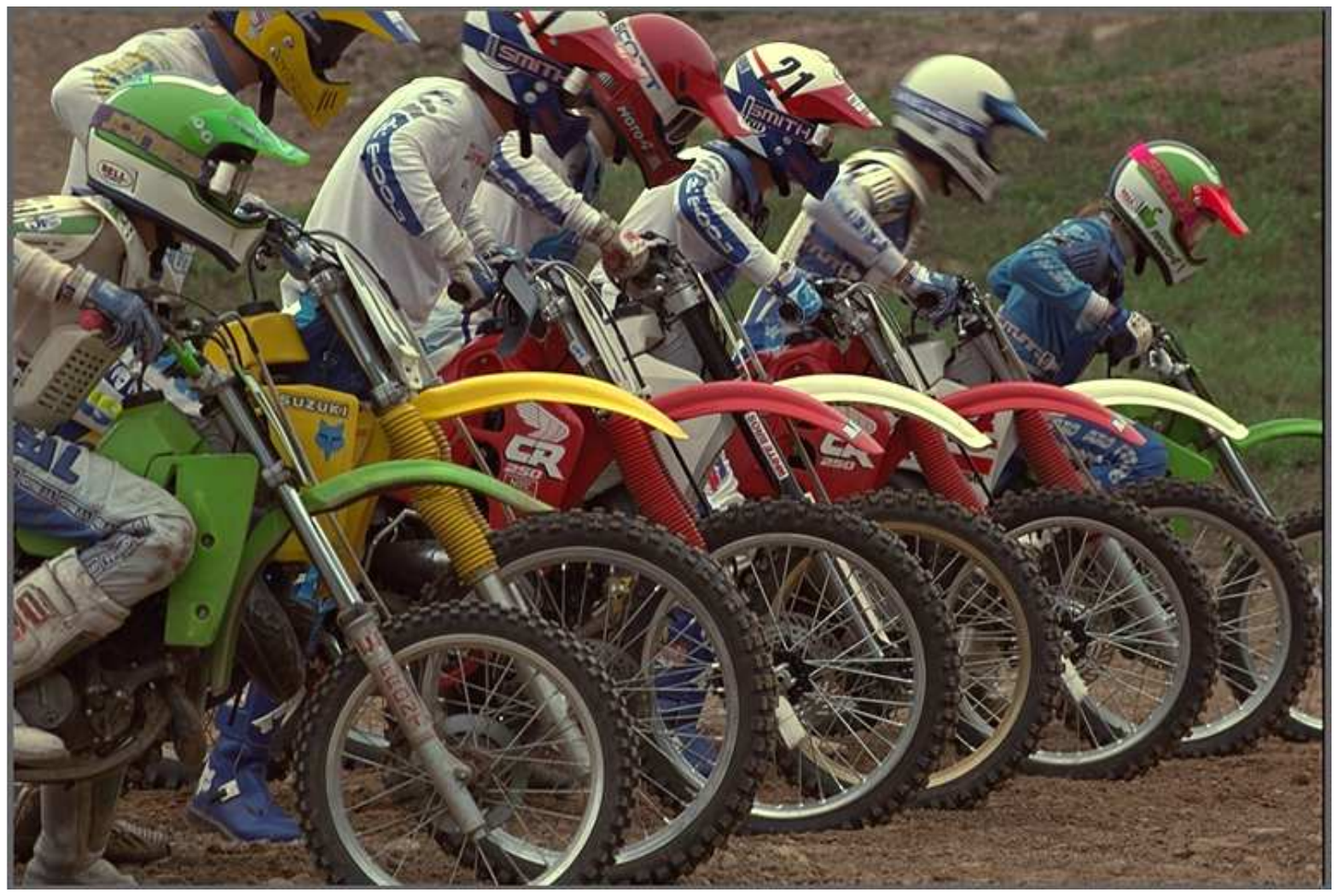}
  \vspace{0.03cm}
  \centerline{\footnotesize{(a)Reference Image}}
\end{minipage}
  \vspace{0.2cm}
\hfill
\begin{minipage}[b]{.48\linewidth}
  \centering
\includegraphics[width=0.60\linewidth, trim= 25mm 90mm 25mm 90mm]{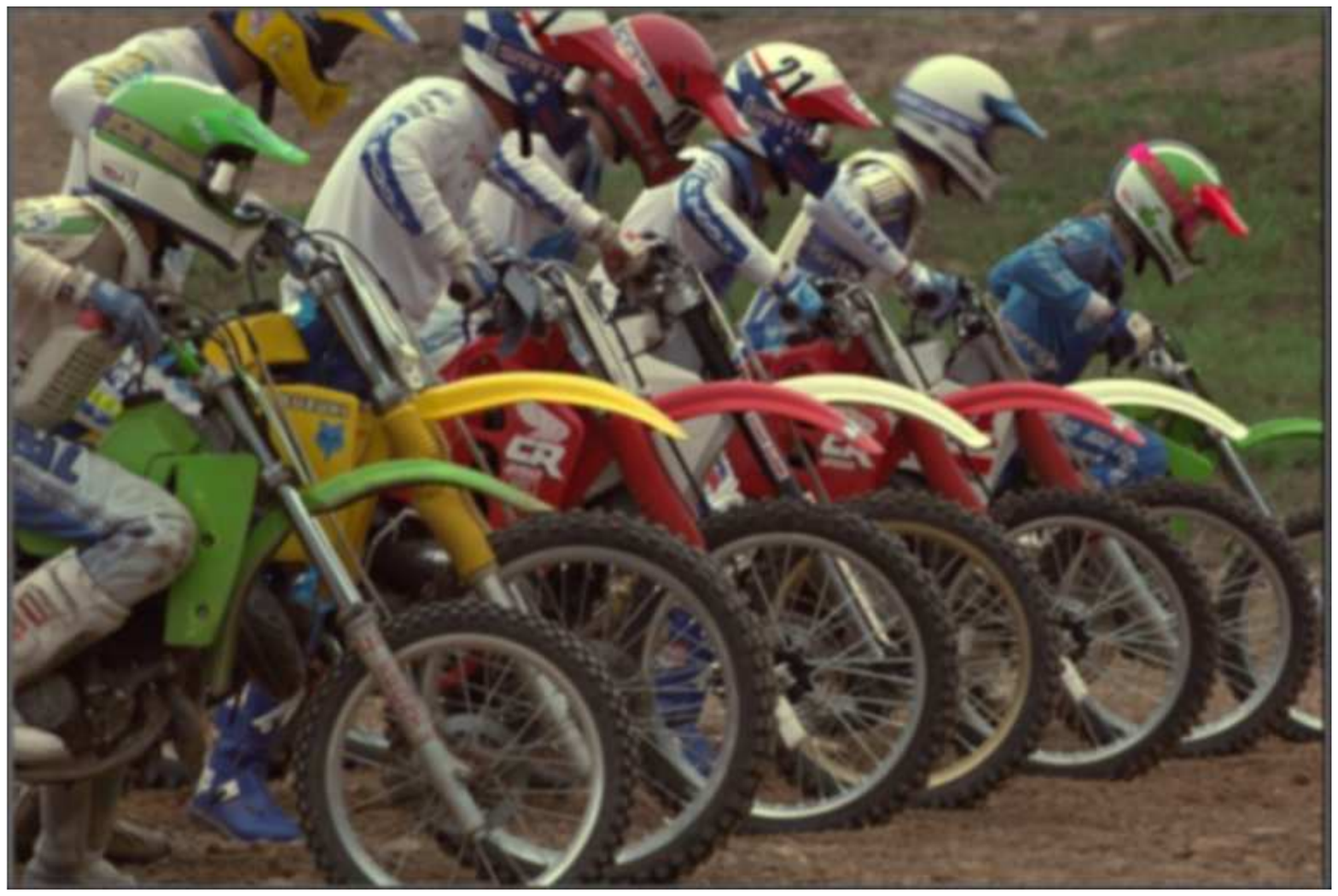}
  \vspace{0.03cm}
  \centerline{\footnotesize{(b) Distorted Image }}
\end{minipage}

\begin{minipage}[b]{.48\linewidth}
  \centering
\includegraphics[width=0.60\linewidth, trim= 25mm 80mm 25mm 80mm]{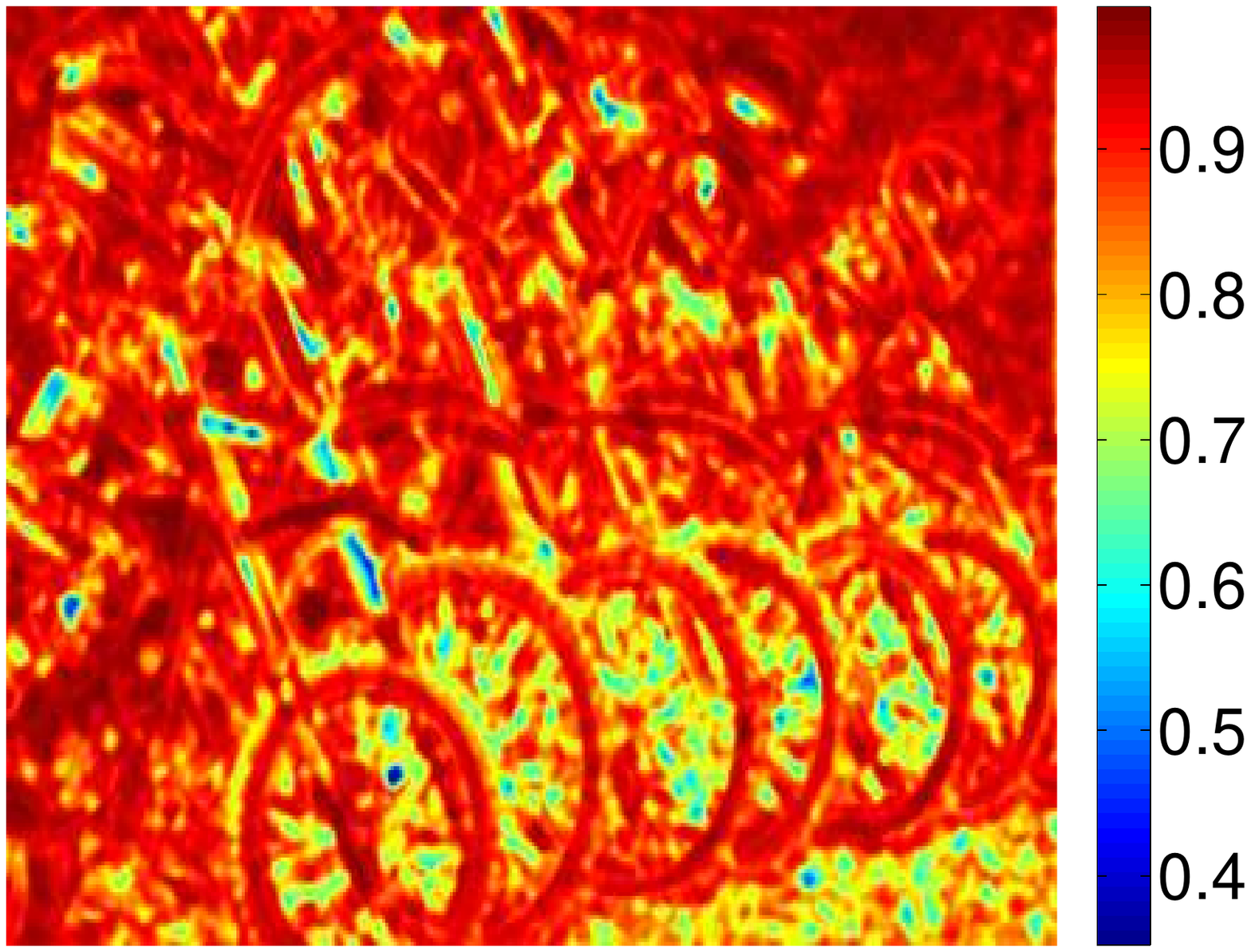}
  \vspace{0.03 cm}
  \centerline{\footnotesize{(c) SSIM Map } }
\end{minipage}
  \vspace{0.2cm}
\hfill
\begin{minipage}[b]{0.48\linewidth}
  \centering
\includegraphics[width=0.60\linewidth, trim= 25mm 80mm 25mm 80mm]{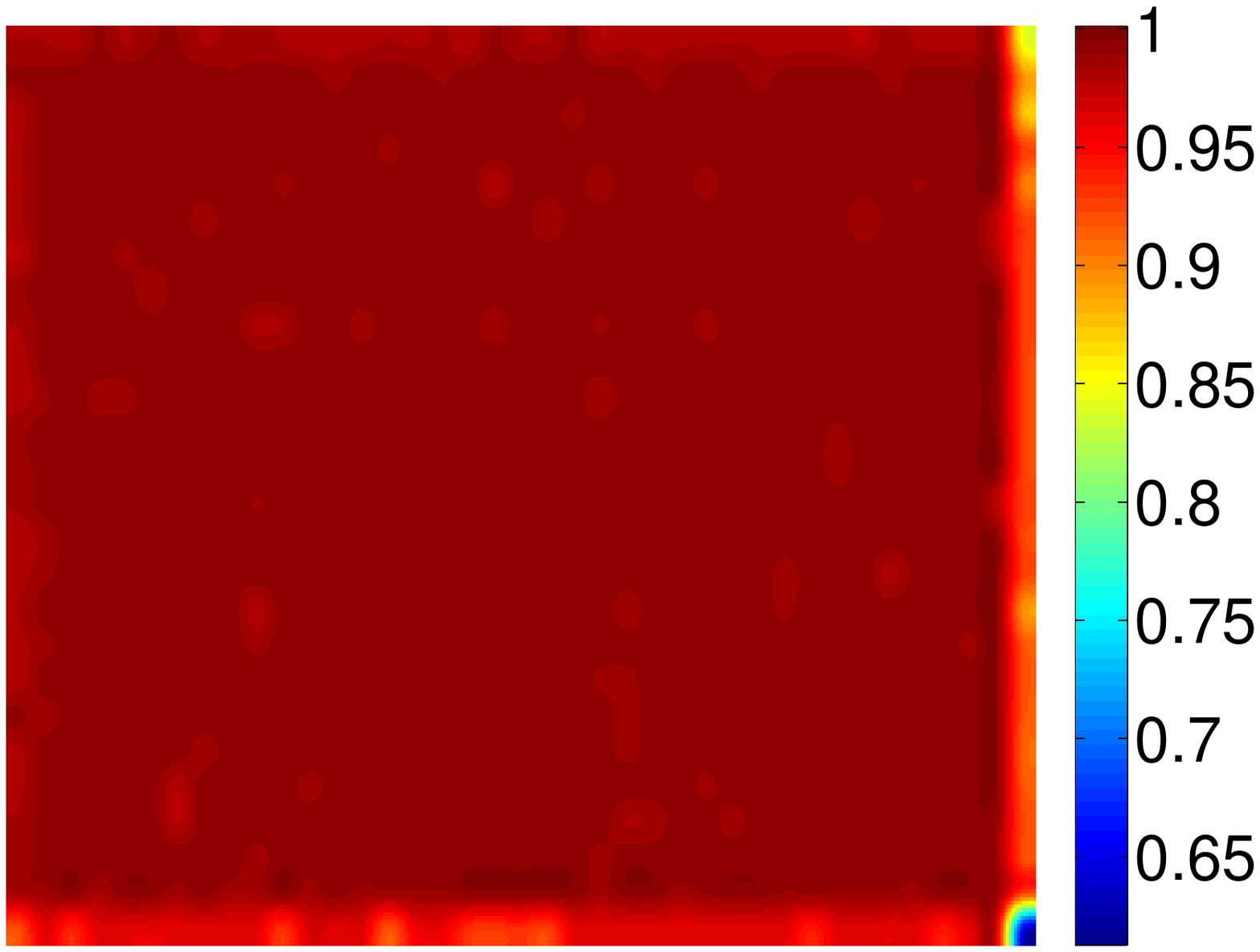}
  \vspace{0.03 cm}
  \centerline{\footnotesize{(d)  Res. PCDM Map with SR=0.05}}
\end{minipage}

\begin{minipage}[b]{.48\linewidth}
  \centering
\includegraphics[width=0.60\linewidth, trim= 25mm 80mm 25mm 80mm]{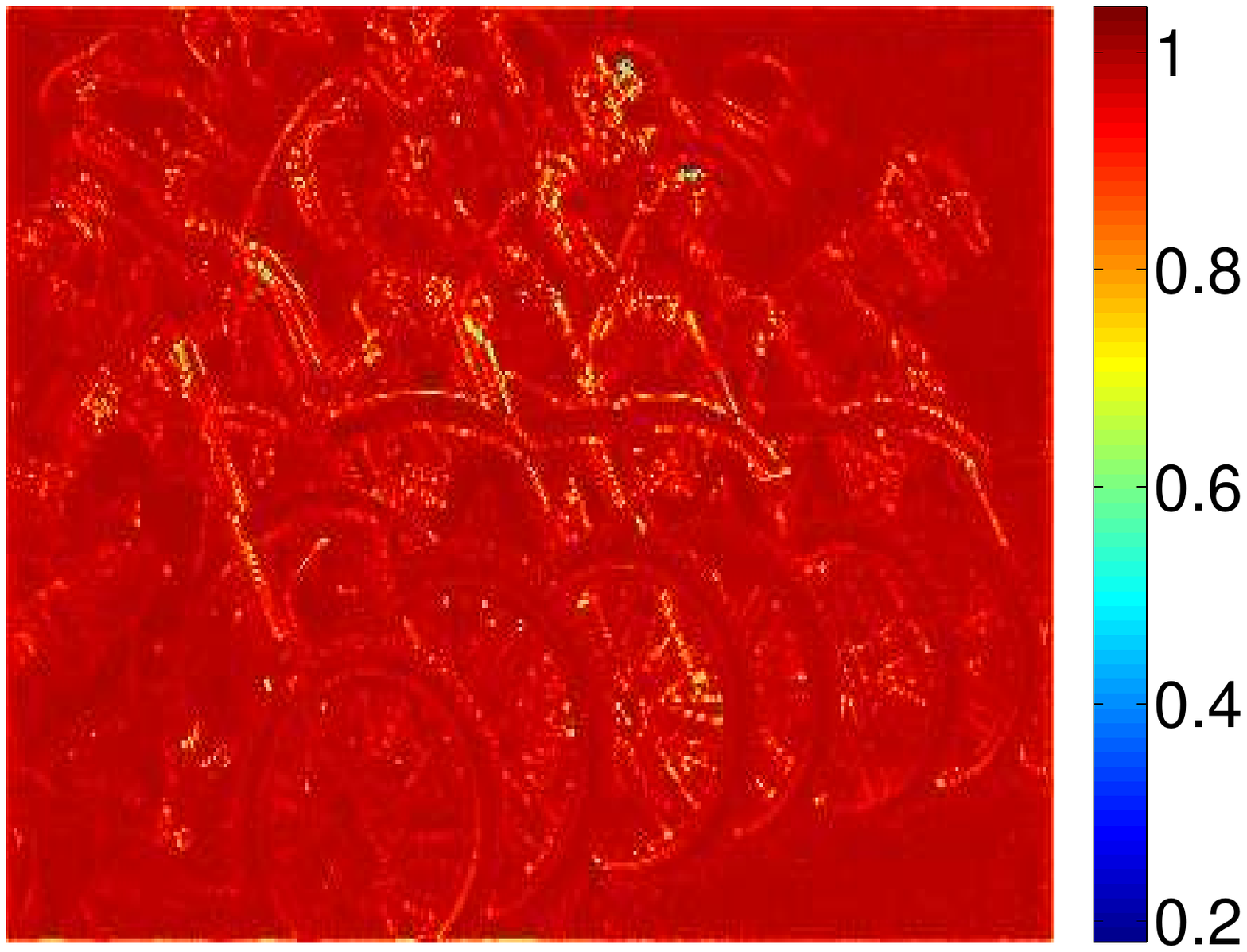}
  \vspace{0.03 cm}
  \centerline{\footnotesize{(e) Res. PCDM Map with SR=0.50 } }
\end{minipage}
  \vspace{0.2cm}
\hfill
\begin{minipage}[b]{0.48\linewidth}
  \centering
\includegraphics[width=0.60\linewidth, trim= 25mm 80mm 25mm 80mm]{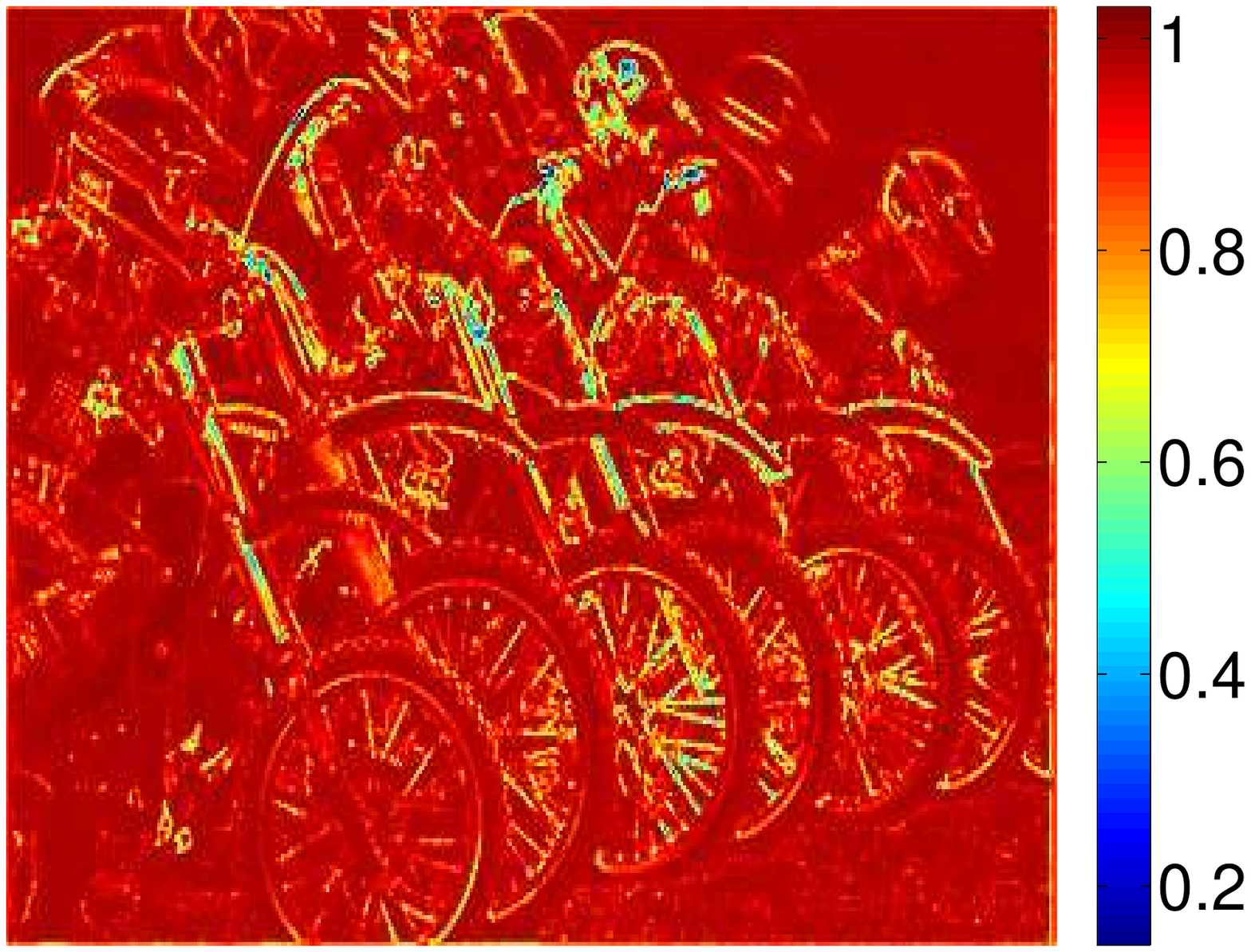}
  \vspace{0.03 cm}
  \centerline{\footnotesize{(f) Res. PCDM Map with SR=1.00}}
\end{minipage}

\vspace{-.3cm}
\caption{Reference image, distorted image and corresponding distortion maps. SR: sampling rate, Res:Residual}
  \vspace{-0.3 cm}
\label{fig:gaussianblurmaps}
\end{figure}

We randomly selected the \textsf{bikes} image from the LIVE database to visualize the distortion maps. However, the same observations are also valid for other images. Reference and distorted images are shown in Fig. \ref{fig:gaussianblurmaps}. Circular-symmetric 2-D Gaussian kernel with a standard deviation of 1.3 is used to degrade the reference image. In general, overall quality of the images is calculated by taking the average over the distortion maps. However, in Fig. \ref{fig:gaussianblurmaps}, we directly show the distortion maps calculated by SSIM and \verb"PCDM" to examine how the metrics perform. Residual of \verb"PCDM" is plotted for a fair comparison since high \verb"PCDM" corresponds to low quality and vice versa.

SSIM detects the observable degradations which are mostly around the textured regions as shown in Fig. \ref{fig:gaussianblurmaps}(c).  We examine the effect of downsampling by varying the sampling rate in \verb"PCDM" model from $0.05$ to $0.50$ and then to $1.00$. A sampling rate of $0.05$ results in oversampling of the image and \verb"PCDM" does not work well as shown in Fig. \ref{fig:gaussianblurmaps}(d). When we increase the sampling rate to $0.5$ and to $1.0$, \verb"PCDM" captures the degradations around the textured regions as shown in Fig. \ref{fig:gaussianblurmaps}(e)-(f). We can obtain higher CC and lower RMSE with a higher sampling rate. However, the proposed method becomes infeasible because of the time-complexity. We summarize the changes in the average residual \verb"PCDM"  and the execution time per single image under varying sampling rate in Table \ref{tab:pcdmScaling}. We used an  Intel(R) Core(TM) i7-3770 CPU @3.50 GHz with 32.0 GB ram.

\begin{table}[htbp!]
  \vspace{-0.4 cm}

\begin{center}
\caption{PCDM under varying sampling rate} \label{tab:pcdmScaling}
\begin{tabular}{|c||c|c|}
 \hline
 \textbf{PCDM Sampling Rate}  & \textbf{Avg. Res. PCDM}  & \textbf{Execution Time (sec)} \\
  \hline
 \textbf{0.05} & 0.98 & 1.3\\
 \hline
  \textbf{0.50} & 0.96 & 120.5\\
 \hline
  \textbf{1.00} & 0.93 & 479.5\\
 \hline

\end{tabular}
\end{center}
  \vspace{-0.6 cm}
\end{table}

\begin{figure}[htbp!]
\begin{minipage}[b]{0.48\linewidth}
  \centering
\includegraphics[width=0.60\linewidth, trim= 25mm 90mm 25mm 90mm]{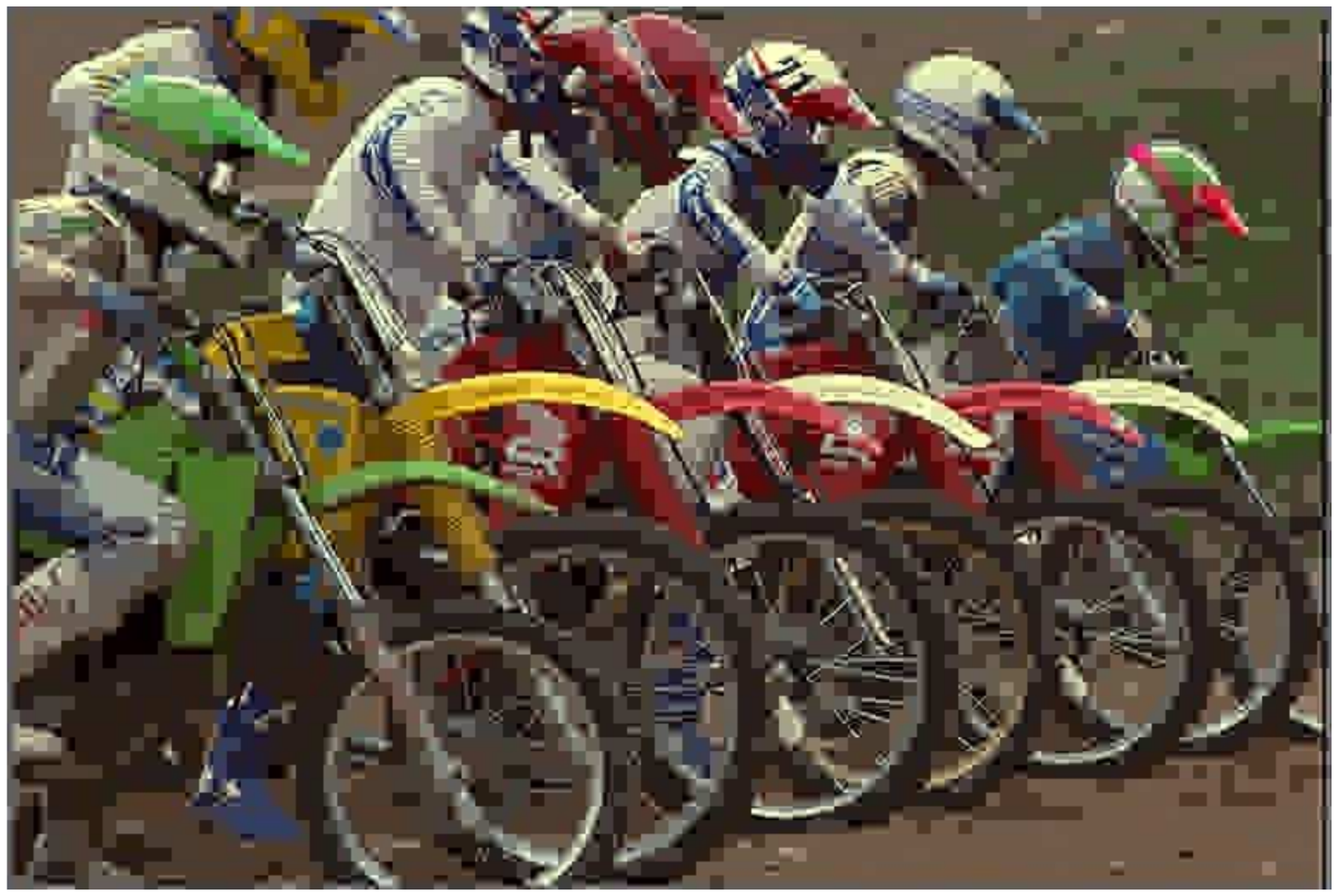}
  \vspace{0.03cm}
  \centerline{\footnotesize{(a)Intensity Distortion }}
\end{minipage}
  \vspace{0.20cm}
\hfill
\begin{minipage}[b]{.48\linewidth}
  \centering
\includegraphics[width=0.60\linewidth, trim= 25mm 90mm 25mm 90mm]{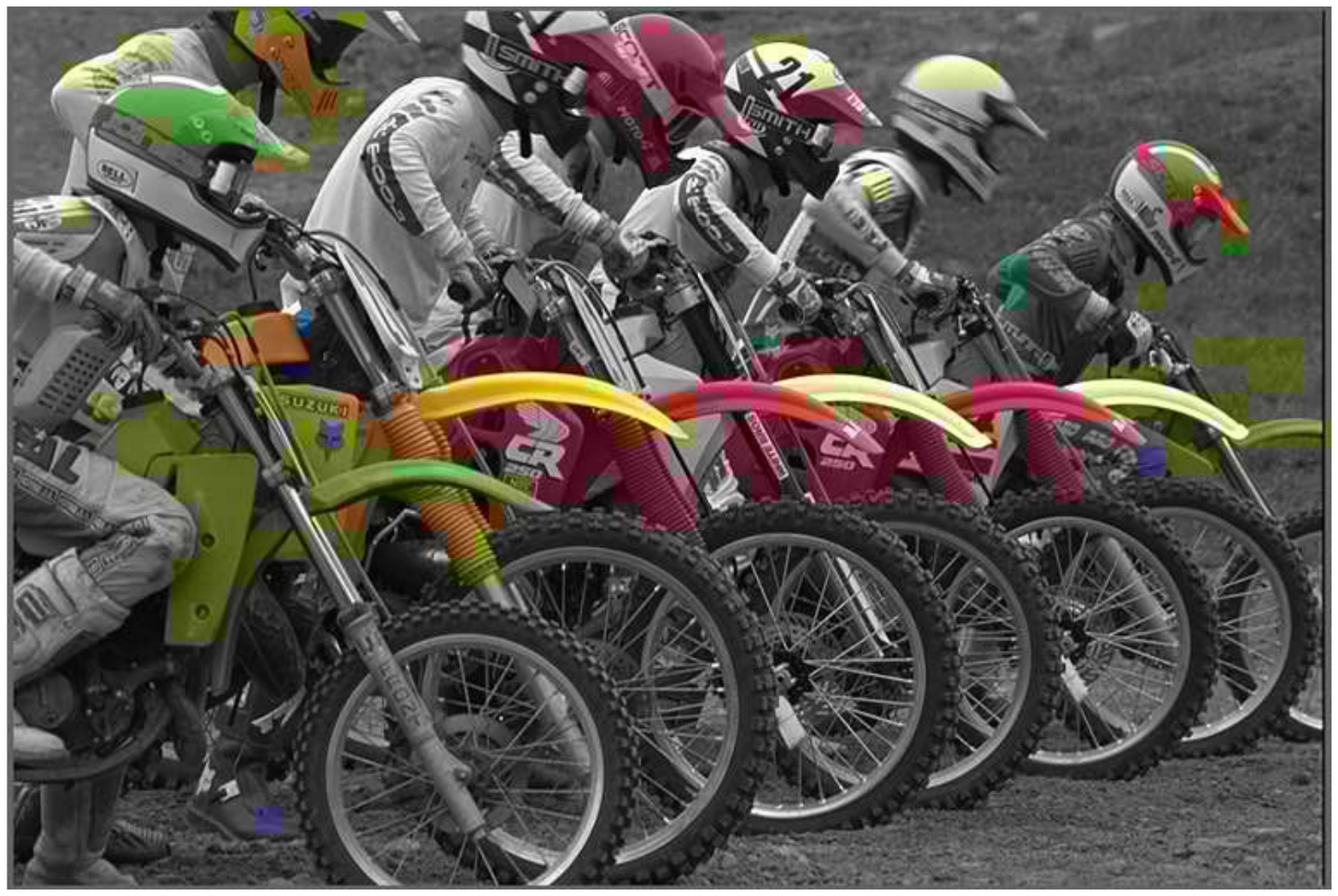}
  \vspace{0.03cm}
  \centerline{\footnotesize{(b) Chroma Distortion }}
\end{minipage}

\begin{minipage}[b]{.48\linewidth}
  \centering
\includegraphics[width=0.60\linewidth, trim= 25mm 80mm 25mm 80mm]{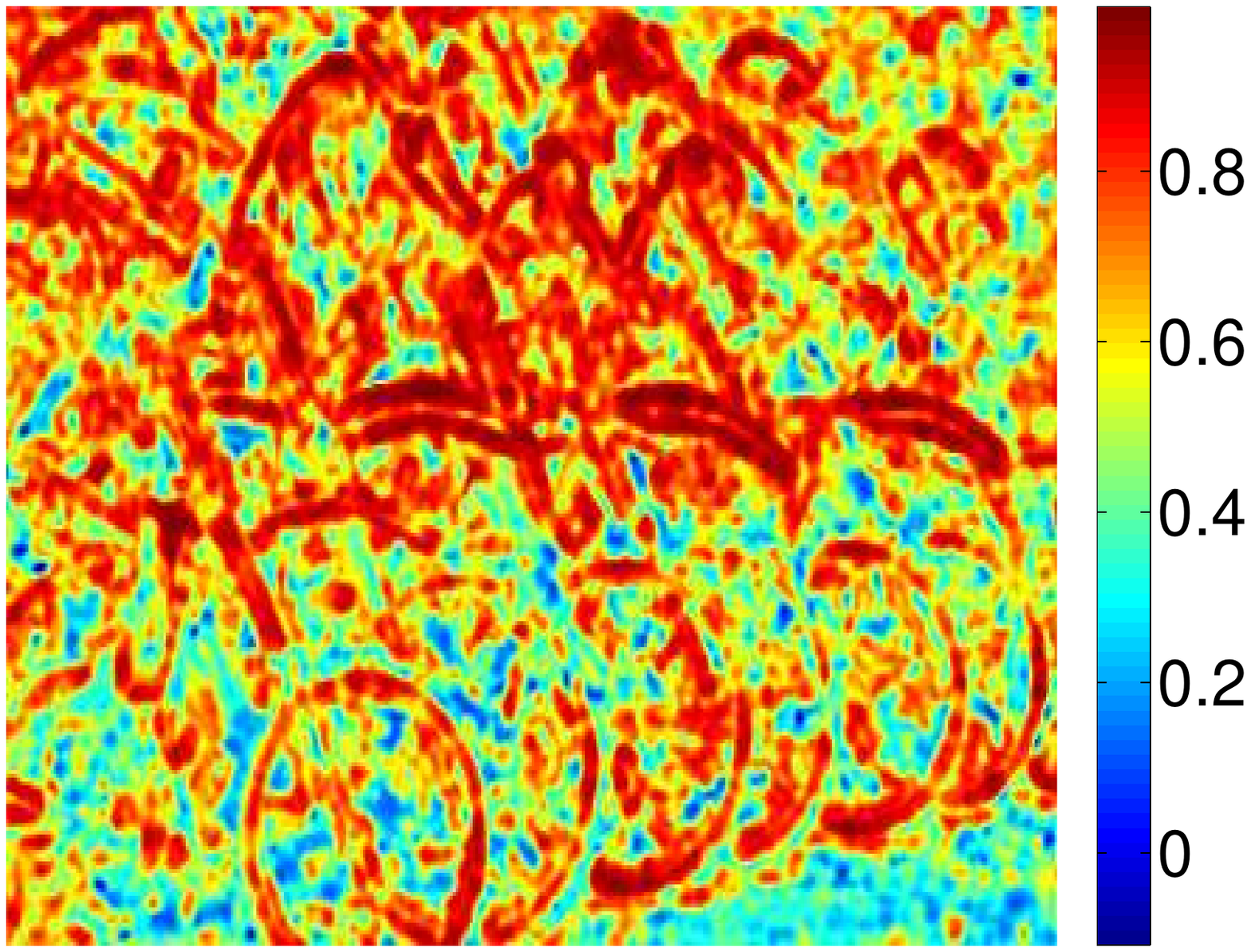}
  \vspace{0.03 cm}
  \centerline{\footnotesize{(c) SSIM Map - Intensity Dist.  } }
\end{minipage}
  \vspace{0.20cm}
\hfill
\begin{minipage}[b]{0.48\linewidth}
  \centering
\includegraphics[width=0.60\linewidth, trim= 25mm 80mm 25mm 80mm]{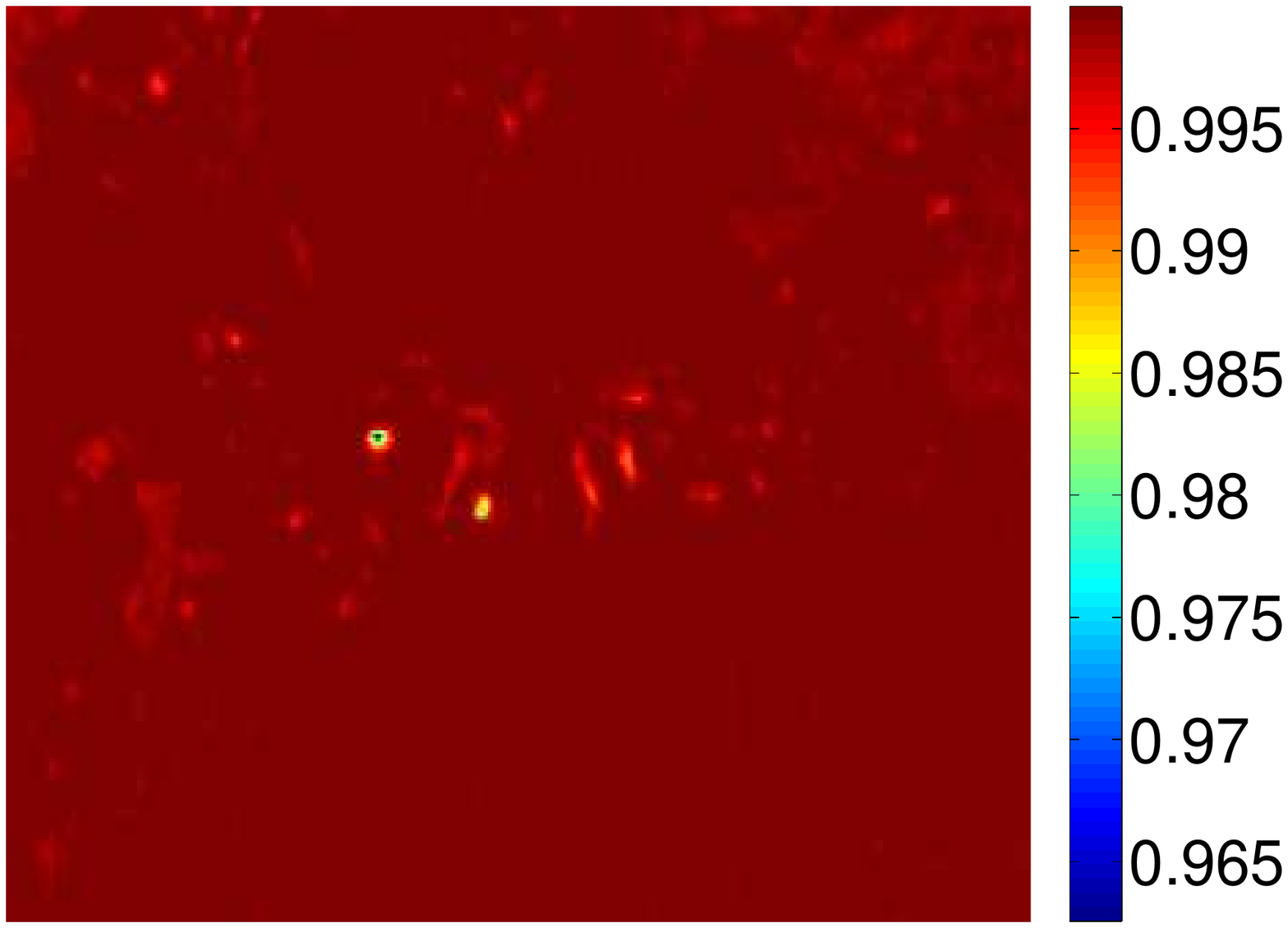}
  \vspace{0.03 cm}
  \centerline{\footnotesize{(d) SSIM Map - Chroma Dist.}}
\end{minipage}

\begin{minipage}[b]{.48\linewidth}
  \centering
\includegraphics[width=0.60\linewidth, trim= 25mm 80mm 25mm 80mm]{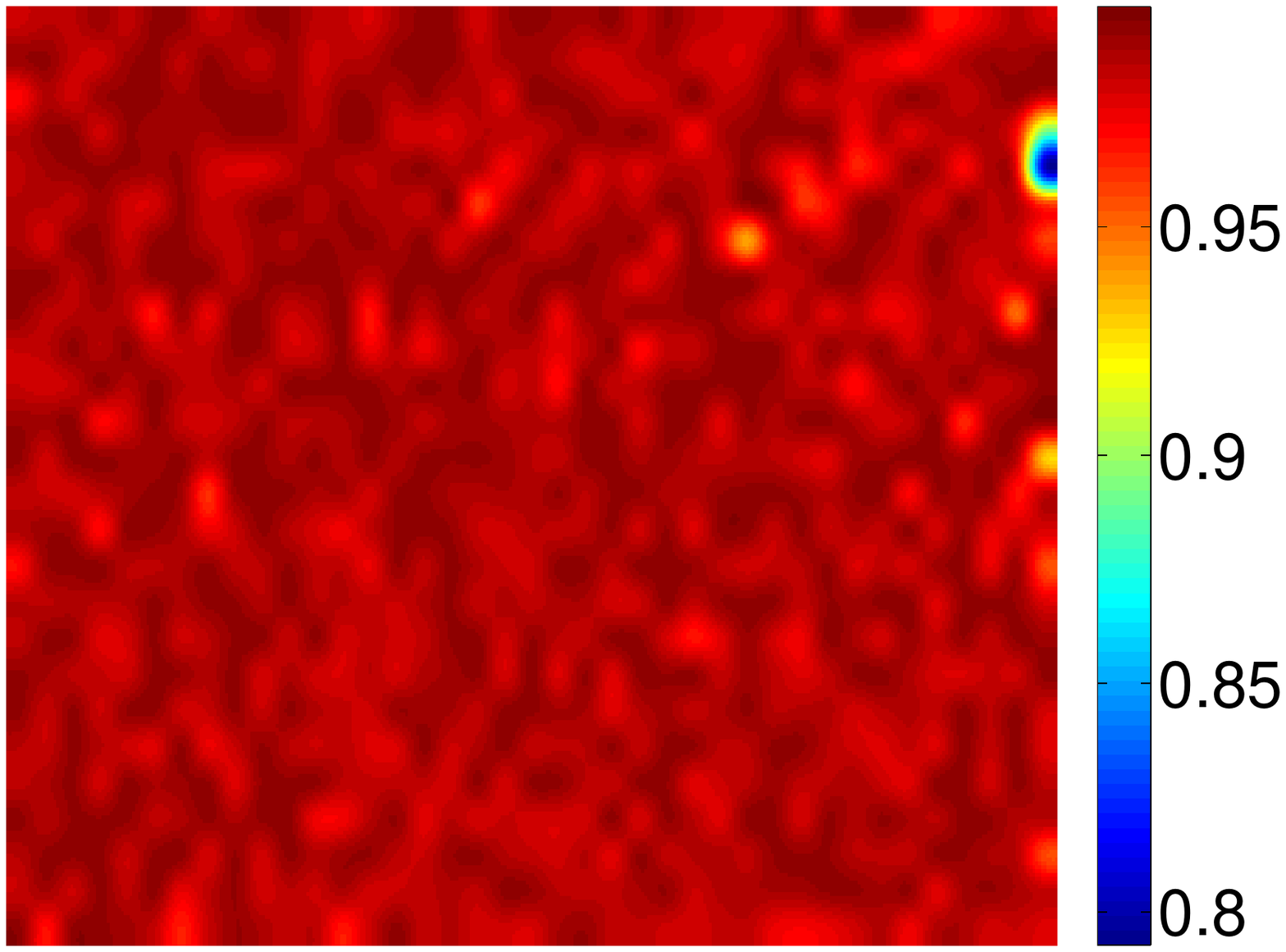}
  \vspace{0.03 cm}
  \centerline{\footnotesize{(e) Res. PCDM Map - Intensity Dist.  } }
\end{minipage}
  \vspace{0.2cm}
\hfill
\begin{minipage}[b]{0.5\linewidth}
  \centering
\includegraphics[width=0.60\linewidth, trim= 25mm 80mm 25mm 80mm]{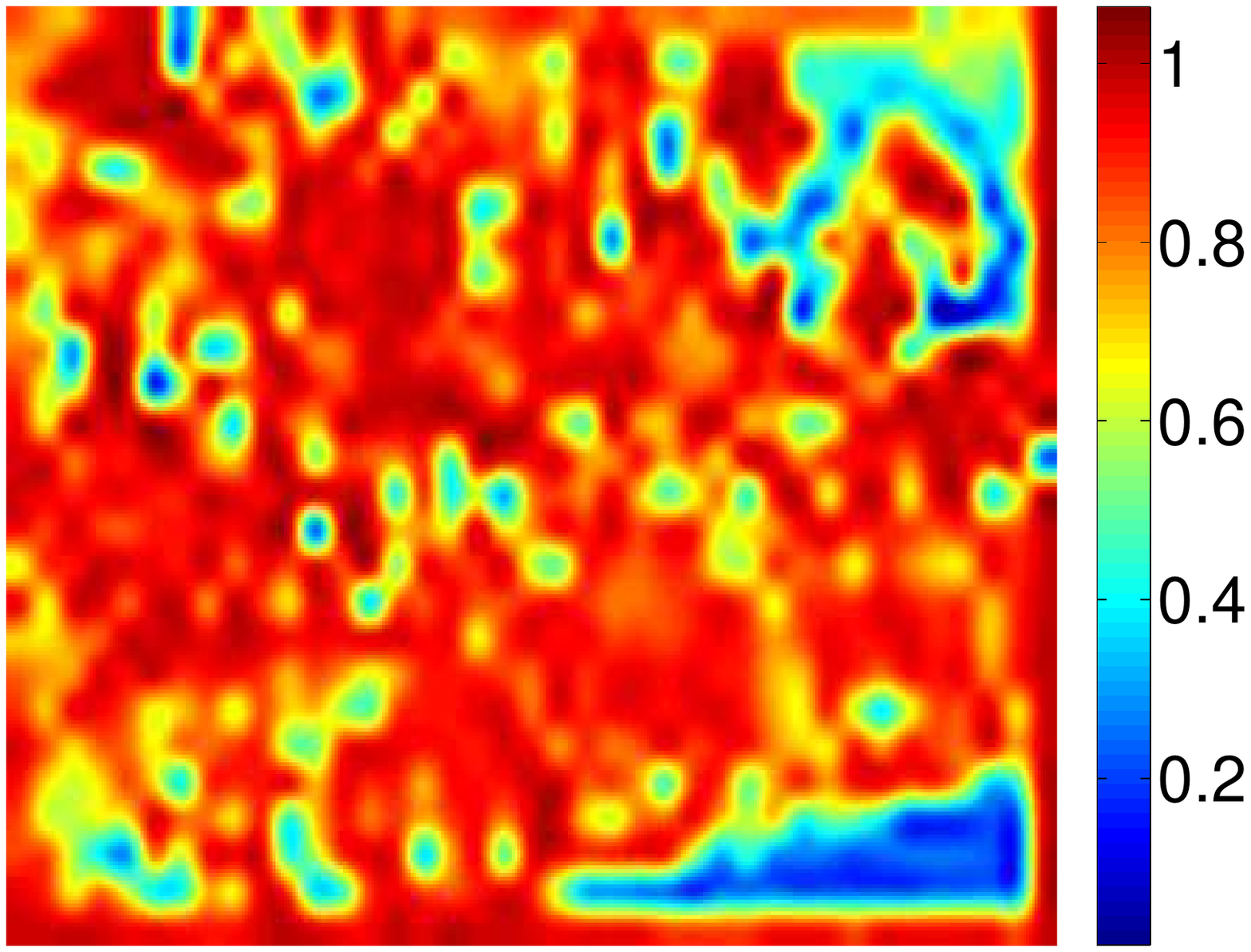}
  \vspace{0.03 cm}
  \centerline{\footnotesize{(f) Res. PCDM Map - Chroma Dist.}}
\end{minipage}

\vspace{-.5cm}
\caption{Intensity and chroma distortion with corresponding distortion maps}\vspace{-.5cm}
\label{fig:jpegmaps}
\end{figure}

Quality metrics such as SSIM, MS-SSIM and CW-SSIM estimate the perceived quality based on the structural cues by neglecting $chroma$ information. However, \verb"PCDM" utilizes both $luma$ and $chroma$. In order to examine how metrics perform under structural and color artifacts,  Jpeg compressed image with a bpp of 0.208 is used. In the visualization, compression artifacts are preferred over Gaussian Blur and White Noise because compression corresponds to a more realistic degradation scenario. The distorted image is converted from the \texttt{RGB} color space to the \texttt{YCbCr} color space. We replace the chrominance channels (\texttt{Cb} and \texttt{Cr}) in the distorted image with the reference to obtain $Intensity$ distortion as given in Fig. \ref{fig:jpegmaps}(a) and we replace the distorted luminance (\texttt{Y}) channel with the error-free channel in the reference to obtain the $Chroma$ distortion as given in Fig. \ref{fig:jpegmaps}(b). $Chroma$ distortion results in loss of color information around some connected regions such as the soil surface and the grass region. We also observe tonal changes over some objects and small regions. Whereas, $Intensity$ distortion leads to blockiness artifacts all over the image especially around textured regions. SSIM detects the blockiness in  $Intensity$ distortion whereas \verb"PCDM" does not capture the distortions with a sampling rate of $0.05$. However, \verb"PCDM" detects color losses at the background region and the tonal changes around foreground objects in the $Chroma$ distortion, which are overlooked by SSIM.

  \vspace{-0.2 cm}

\section{Conclusion}
\vspace{-0.20cm}
\label{sec:conclusion}
In this paper, we extended the CIEDE2000 formula using a perceptual color difference metric to estimate the subjective quality of images. We have shown that perceptual color difference metric results highly correlate with DMOS and it performs better than the  pixel-wise fidelity metrics and the CIEDE2000 formula in terms of correlation and root-mean-square error. However, structural metrics perform better than color-based metric under Fast Fading and Gaussian Blur due to the oversampling in \verb"PCDM". When the sampling ratio is increased, \verb"PCDM" performs better but it also increases the time-complexity significantly. We started combining color and structure based metrics to estimate the quality of experience for the end user and the hybrid metric already leads to promising results in LIVE and TID2013 image databases.  
\end{document}